\documentclass[manyauthors,nocleardouble,COMPASS]{cernphprep}
\usepackage{times}
\usepackage{graphicx}
\usepackage{float}
\usepackage{amsmath}
\usepackage{amssymb}
\usepackage{xcolor}
\usepackage[displaymath,mathlines]{lineno}
\usepackage{booktabs}
\usepackage{hyperref}
\usepackage[normalem]{ulem}
\usepackage{hepunits}

\newcommand{\val}[2]{#1\,#2}  % \val{10}{m} -> 10\,\mathrm{m}
\newcommand{\diff}{\mathrm{d}}  % alias of \mathrm{d} for differenctial.

\newcommand{\secref}[1]{Sec.~\ref{#1}}
\newcommand{\figref}[1]{Fig.~\ref{#1}}

\renewcommand{\eqref}[1]{Eq.~\ref{#1}}

\hypersetup{
     bookmarks=true,         % show bookmarks bar?
     unicode=false,          % non-Latin characters in Acrobat's bookmarks
     pdftoolbar=true,        % show Acrobat's toolbar?
     pdfmenubar=true,        % show Acrobat's menu?
     pdffitwindow=false,     % window fit to page when opened
     pdfstartview={FitH},    % fits the width of the page to the window
     pdfauthor={The COMPASS Collaboration},     % author
     pdfsubject={PWA},   % subject of the document
     pdfnewwindow=true,      % links in new window
     colorlinks=false,       % false: boxed links; true: colored links
     linkcolor=red,          % color of internal links
     citecolor=green,        % color of links to bibliography
     filecolor=magenta,      % color of file links
     urlcolor=cyan           % color of external links
}

\makeatletter
\let\start@align@nopar\start@align
\let\start@gather@nopar\start@gather
\let\start@multline@nopar\start@multline
\long\def\start@align{\par\start@align@nopar}
\long\def\start@gather{\par\start@gather@nopar}
\long\def\start@multline{\par\start@multline@nopar}
\makeatother
\begin{document}
\begin{titlepage}
\PHnumber{2015-245}
\PHdate{\begin{tabular}[c]{r@{}}
September 11, 2015\\
rev. Dec. 14, 2015
\end{tabular}
\vspace*{-0.6cm}}
\vspace{1cm}
\title{\LARGE
 \textbf{Longitudinal double spin asymmetries in single hadron quasi-real photoproduction at high \boldmath$p_T$}}
\vspace*{0.5cm}
%\Collaboration{The COMPASS Collaboration}
%\ShortAuthor{The COMPASS Collaboration}
%\begin{frontmatter}

%\title{
% Longitudinal double spin asymmetries in single hadron quasi-real photoproduction at high $p_T$}
%\input{PLBauthors}

\begin{abstract}
We measured the longitudinal double spin asymmetries $A_{LL}$ for single hadron
muoproduction off protons and deuterons at photon virtuality $Q^2 < 1
(\GeVoverc)^2$ for transverse hadron momenta $p_T$ in the range $1$~\GeVoverc\ 
to $4$~\GeVoverc . They were determined using COMPASS data
taken with a polarised muon beam of $160$~\GeVoverc \ or $200$~\GeVoverc \ 
impinging on polarised $\mathrm{{}^6LiD}$ or $\mathrm{NH_3}$ targets. The
experimental asymmetries are compared to next-to-leading order pQCD
calculations, and are  sensitive to the gluon polarisation $\Delta G$ inside
the nucleon in the range of the nucleon momentum fraction carried by gluons
$0.05 < x_g < 0.2$.
%\end{abstract}
%\end{frontmatter}

\vspace{3em}
%{\it key words:}
%COMPASS,
%Deep inelastic scattering,
%Double spin asymmetry,
%high $p_T$,
%$\Delta G$
%\\[3em]
\end{abstract}

\vfill
\Submitted{(Submitted to Physics Letters B)}
\end{titlepage}
{\pagestyle{empty} }
%%%%%%%%%%%%%%%%%%%%%%%%%%%%%%%%%%%%%%%%%%%%%%%%%%%%%%%%%%%%%%%
%
% 2015_auththorlist.tex  
%
%%%%%%%%%%%%%%%%%%%%%%%%%%%%%%%%%%%%%%%%%%%%%%%%%%%%%%%%%%%%%%%
\section*{The COMPASS Collaboration}
\label{app:collab}
\renewcommand\labelenumi{\textsuperscript{\theenumi}~}
\renewcommand\theenumi{\arabic{enumi}}
\begin{flushleft}
C.~Adolph\Irefn{erlangen},
R.~Akhunzyanov\Irefn{dubna}, %phd
M.G.~Alexeev\Irefn{turin_u},
G.D.~Alexeev\Irefn{dubna}, %1
A.~Amoroso\Irefnn{turin_u}{turin_i},
V.~Andrieux\Irefn{saclay},
V.~Anosov\Irefn{dubna}, %2
W.~Augustyniak\Irefn{warsaw},
A.~Austregesilo\Irefn{munichtu},
C.D.R.~Azevedo\Irefn{aveiro}
% ...
\end{flushleft}
%%%%%%%%%%%%%%%%%%%%%%%%%%%%%%%%%%%%%%%%%%%%%%%%%%%%%%%%%%%%%%%%%%%%%%%%%%%%%%%%%%%%%%%%%%%%%%%%%%%%%%%%%%%%%%%%%%%%%%%
%
% institutes
%
%%%%%%%%%%%%%%%%%%%%%%%%%%%%%%%%%%%%%%%%%%%%%%%%%%%%%%%%%%%%%%%%%%%%%%%%%%%%%%%%%%%%%%%%%%%%%%%%%%%%%%%%%%%%%%%%%%%%%%%
%\item \Idef{bielefeld}{Universit\"at Bielefeld, Fakult\"at f\"ur Physik, 33501 Bielefeld, Germany\Arefs{l}}
%\item \Idef{munichlmu}{Ludwig-Maximilians-Universit\"at M\"unchen, Department f\"ur Physik, 80799 Munich, Germany\Arefs{l}\Arefs{r}}
\begin{Authlist}
\item \Idef{turin_p}{University of Eastern Piedmont, 15100 Alessandria, Italy}
\item \Idef{aveiro}{University of Aveiro, Department of Physics, 3810-193 Aveiro, Portugal} 
\item \Idef{bochum}{Universit\"at Bochum, Institut f\"ur Experimentalphysik, 44780 Bochum, Germany\Arefs{l}\Arefs{s}}
\item \Idef{bonniskp}{Universit\"at Bonn, Helmholtz-Institut f\"ur  Strahlen- und Kernphysik, 53115 Bonn, Germany\Arefs{l}}
\item \Idef{bonnpi}{Universit\"at Bonn, Physikalisches Institut, 53115 Bonn, Germany\Arefs{l}}
\item \Idef{brno}{Institute of Scientific Instruments, AS CR, 61264 Brno, Czech Republic\Arefs{m}}
\item \Idef{calcutta}{Matrivani Institute of Experimental Research \& Education, Calcutta-700 030, India\Arefs{n}}
\item \Idef{dubna}{Joint Institute for Nuclear Research, 141980 Dubna, Moscow region, Russia\Arefs{o}}
\item \Idef{erlangen}{Universit\"at Erlangen--N\"urnberg, Physikalisches Institut, 91054 Erlangen, Germany\Arefs{l}}
\item \Idef{freiburg}{Universit\"at Freiburg, Physikalisches Institut, 79104 Freiburg, Germany\Arefs{l}\Arefs{s}}
\item \Idef{cern}{CERN, 1211 Geneva 23, Switzerland}
\item \Idef{liberec}{Technical University in Liberec, 46117 Liberec, Czech Republic\Arefs{m}}
\item \Idef{lisbon}{LIP, 1000-149 Lisbon, Portugal\Arefs{p}}
\item \Idef{mainz}{Universit\"at Mainz, Institut f\"ur Kernphysik, 55099 Mainz, Germany\Arefs{l}}
\item \Idef{miyazaki}{University of Miyazaki, Miyazaki 889-2192, Japan\Arefs{q}}
\item \Idef{moscowlpi}{Lebedev Physical Institute, 119991 Moscow, Russia}
\item \Idef{munichtu}{Technische Universit\"at M\"unchen, Physik Department, 85748 Garching, Germany\Arefs{l}\Arefs{r}}
\item \Idef{nagoya}{Nagoya University, 464 Nagoya, Japan\Arefs{q}}
\item \Idef{praguecu}{Charles University in Prague, Faculty of Mathematics and Physics, 18000 Prague, Czech Republic\Arefs{m}}
\item \Idef{praguectu}{Czech Technical University in Prague, 16636 Prague, Czech Republic\Arefs{m}}
\item \Idef{protvino}{State Scientific Center Institute for High Energy Physics of National Research Center `Kurchatov Institute', 142281 Protvino, Russia}
\item \Idef{saclay}{CEA IRFU/SPhN Saclay, 91191 Gif-sur-Yvette, France\Arefs{s}}
\item \Idef{taipei}{Academia Sinica, Institute of Physics, Taipei, 11529 Taiwan}
\item \Idef{telaviv}{Tel Aviv University, School of Physics and Astronomy, 69978 Tel Aviv, Israel\Arefs{t}}
\item \Idef{triest_u}{University of Trieste, Department of Physics, 34127 Trieste, Italy}
\item \Idef{triest_i}{Trieste Section of INFN, 34127 Trieste, Italy}
\item \Idef{triest_ictp}{Abdus Salam ICTP, 34151 Trieste, Italy}
\item \Idef{turin_u}{University of Turin, Department of Physics, 10125 Turin, Italy}
\item \Idef{turin_i}{Torino Section of INFN, 10125 Turin, Italy}
\item \Idef{illinois}{University of Illinois at Urbana-Champaign, Department of Physics, Urbana, IL 61801-3080, U.S.A.}   
\item \Idef{warsaw}{National Centre for Nuclear Research, 00-681 Warsaw, Poland\Arefs{u} }
\item \Idef{warsawu}{University of Warsaw, Faculty of Physics, 02-093 Warsaw, Poland\Arefs{u} }
\item \Idef{warsawtu}{Warsaw University of Technology, Institute of Radioelectronics, 00-665 Warsaw, Poland\Arefs{u} }
\item \Idef{yamagata}{Yamagata University, Yamagata, 992-8510 Japan\Arefs{q} }
\end{Authlist}
%%%%%%%%%%%%%%%%%%%%%%%%%%%%%%%%%%%%%%%%%%%%%%%%%%%%%%%%%%%%%%%%%%%%%%%%%%%%%%%%%%%%%%%%%%%%%%%%%%%%%%%%%%%%%%%%%%%%%%%
%
% Notes
%
%%%%%%%%%%%%%%%%%%%%%%%%%%%%%%%%%%%%%%%%%%%%%%%%%%%%%%%%%%%%%%%%%%%%%%%%%%%%%%%%%%%%%%%%%%%%%%%%%%%%%%%%%%%%%%%%%%%%%%%
%\item \Adef{a0}{Retired from Universit\"at Bielefeld, Fakult\"at f\"ur Physik, 33501 Bielefeld, Germany}
%\item \Adef{a1}{Present address: Universit\"at Mainz, Helmholtz-Institut f\"ur Strahlen- und Kernphysik, 55099 Mainz, Germany}
%\vspace*{-\baselineskip}
\renewcommand\theenumi{\alph{enumi}}
\begin{Authlist}
\item [{\makebox[2mm][l]{\textsuperscript{*}}}] Deceased
\item \Adef{a}{Also at Instituto Superior T\'ecnico, Universidade de Lisboa, Lisbon, Portugal}
\item \Adef{b}{Also at Department of Physics, Pusan National University, Busan 609-735, Republic of Korea and at Physics Department, Brookhaven National Laboratory, Upton, NY 11973, U.S.A. }
\item \Adef{r}{Supported by the DFG cluster of excellence `Origin and Structure of the Universe' (www.universe-cluster.de)}
\item \Adef{d}{Also at Chubu University, Kasugai, Aichi, 487-8501 Japan\Arefs{q}}
\item \Adef{e}{Also at KEK, 1-1 Oho, Tsukuba, Ibaraki, 305-0801 Japan}
\item \Adef{f}{Present address: Universit\"at Bonn, Helmholtz-Institut f\"ur Strahlen- und Kernphysik, 53115 Bonn, Germany}
\item \Adef{g}{Also at Moscow Institute of Physics and Technology, Moscow Region, 141700, Russia}
\item \Adef{j}{Present address: Typesafe AB, Dag Hammarskj\"olds v\"ag 13, 752 37 Uppsala, Sweden}
\item \Adef{v}{Supported by Presidential grant NSh - 999.2014.2}
\item \Adef{z}{Corresponding author}
\item \Adef{h}{Present address: RWTH Aachen University, III. Physikalisches Institut, 52056 Aachen, Germany}
\item \Adef{i}{Present address: Uppsala University, Box 516, SE-75120 Uppsala, Sweden}
\item \Adef{c}{Supported by the DFG Research Training Group Programme 1102  ``Physics at Hadron Accelerators''}
%
% institutes
%
\item \Adef{l}{Supported by the German Bundesministerium f\"ur Bildung und Forschung}
\item \Adef{s}{Supported by EU FP7 (HadronPhysics3, Grant Agreement number 283286)}
\item \Adef{m}{Supported by Czech Republic MEYS Grant LG13031}
\item \Adef{n}{Supported by SAIL (CSR), Govt.\ of India}
\item \Adef{o}{Supported by CERN-RFBR Grant 12-02-91500}
\item \Adef{p}{\raggedright Supported by the Portuguese FCT - Funda\c{c}\~{a}o para a Ci\^{e}ncia e Tecnologia, COMPETE and QREN,
 Grants CERN/FP 109323/2009, 116376/2010, 123600/2011 and CERN/FIS-NUC/0017/2015}
\item \Adef{q}{Supported by the MEXT and the JSPS under the Grants No.18002006, No.20540299 and No.18540281; Daiko Foundation and Yamada Foundation}
\item \Adef{t}{Supported by the Israel Academy of Sciences and Humanities}
\item \Adef{u}{Supported by the Polish NCN Grant DEC-2011/01/M/ST2/02350}
%
% Dünnweber, Faessler, Schlüter for 3-pi paper
%
%
\end{Authlist}

%\newpage
%
%%\linenumbers
%

\section{Introduction}
\label{sec:Introduction}

The spin structure of the nucleon is one of the major unresolved issues in
hadronic physics. While the quark spin contribution to the nucleon spin,
denoted as $\Delta \Sigma$, has been measured to be about 30\%
\cite{RevModPhys.85.655}, the gluon spin contribution is still insufficiently
constrained after more than two decades of intense study. In the framework of
perturbative Quantum Chromodynamics (pQCD), inclusive Deep Inelastic Scattering
(DIS) is sensitive to gluon contributions only through higher-order corrections
to the cross section. The spin-averaged gluon density  $g(x_g)$, where $x_g$
denotes the nucleon momentum fraction carried by gluons, is well constrained by
DIS experiments with unpolarised beam and target because of their high
statistics and large kinematic coverage. The fewer data from DIS experiments
with polarised beam and target, however, can not sufficiently constrain the
gluon helicity distribution $\Delta g(x_g)$. This affects directly our
knowledge of the contribution of the gluon spin to the spin of the nucleon,
known as $\Delta G = \int \Delta g(x_g) dx_g$, and to a lesser extent that of
the quarks~\cite{COMPASS:2015g1p}. In order to better constrain $\Delta
g(x_g)$, one has to resort to processes where contributions from gluons appear
at leading order, such as hadron production at high transverse momenta or
production of open charm in polarised
lepton--nucleon~\cite{SMCdg,Ageev:2005pq,HERMES,Adolph:2012vj,COMPASSoc} or
hadron--hadron interactions~\cite{STAR1,PHENIX1,PHENIX2,STAR2}.\\

The COMPASS collaboration has already investigated asymmetries of hadrons
at high transverse momenta $p_T$, in both the DIS and the quasi-real photoproduction
regimes~\cite{Ageev:2005pq,Adolph:2012vj,Marcin}. Here,
transverse means transverse with respect to the direction of the virtual photon
$\gamma^*$ that is exchanged in the scattering process. Using a Lund Monte
Carlo simulation, these measurements were interpreted on the hadron level,
thereby simultaneously extracting the gluon helicity on the parton level. Such
an analysis is 
restricted to leading  order (LO) in the strong coupling constant $\alpha_s$,
as presently there exists no next-to-leading order (NLO) Monte Carlo simulation
for leptoproduction. Due to the limitation of neglecting gluon contributions at
NLO, such results can not  be used in recent global fits at NLO of polarised
Parton Distribution Functions (PDFs)~\cite{deFlorian:2014yva}.

\begin{figure}[H]
 \begin{minipage}[b]{0.5\linewidth}
  \center
  \includegraphics[width=8cm]{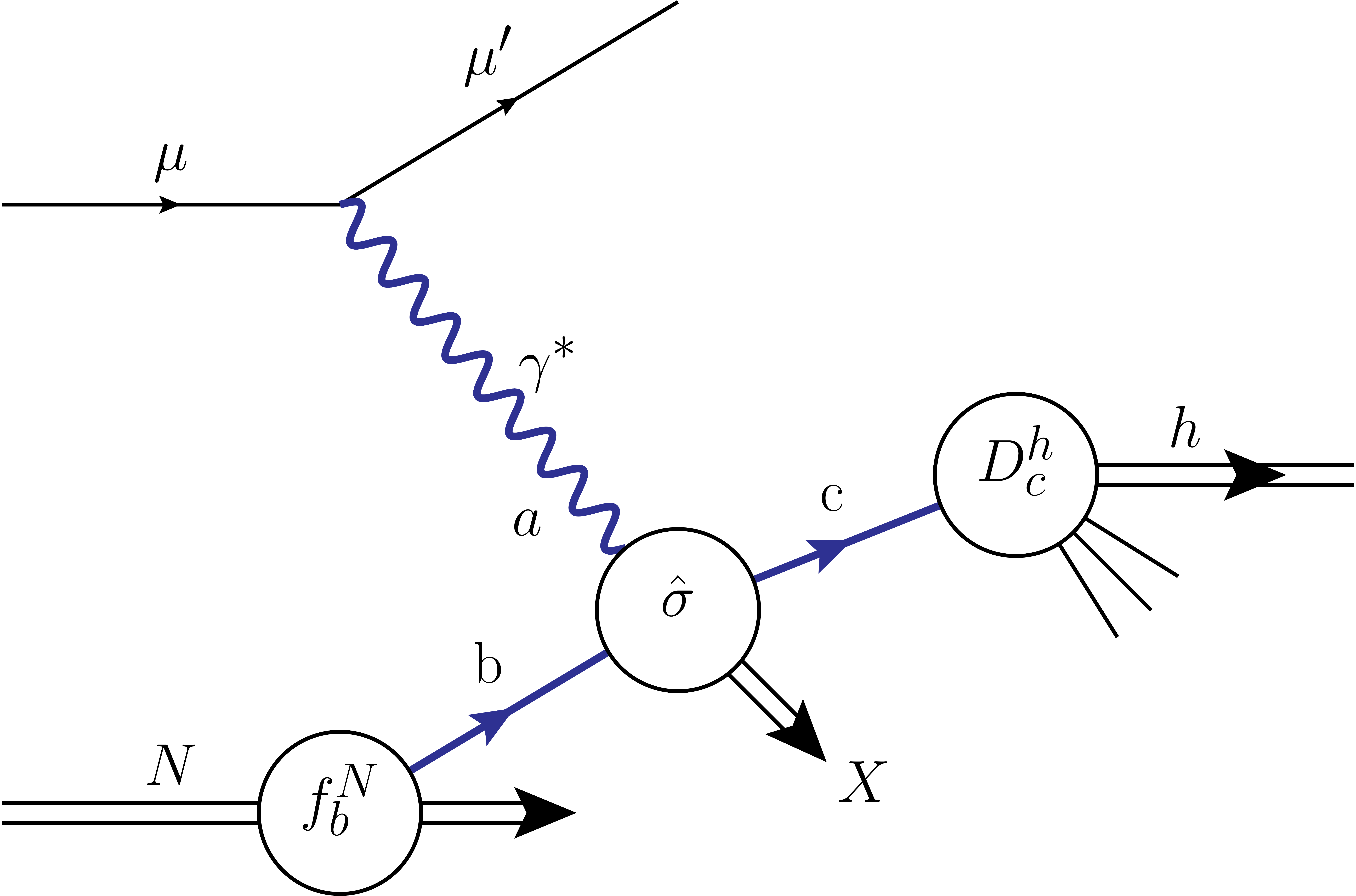}
 \end{minipage}
 \begin{minipage}[b]{0.5\linewidth}
  \center
  \includegraphics[width=8cm]{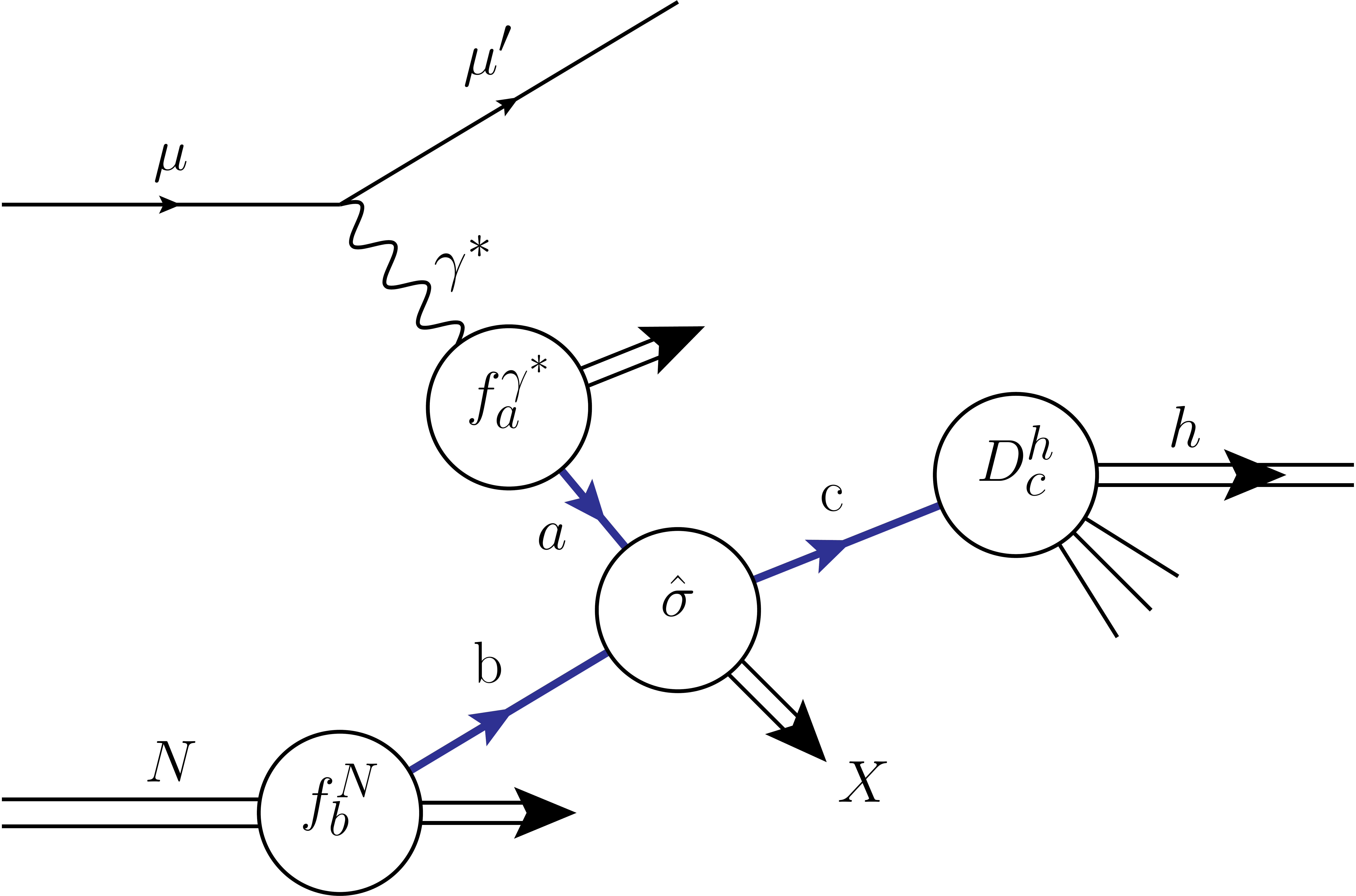}
 \end{minipage}
 \caption{Contributions to the single-inclusive cross section for quasi-real
photoproduction of a hadron $h$ into direct (left) and resolved (right)
subprocesses according to Ref.~\cite{Jager:2005uf}. The internal lines
represent the photon $a=\gamma^*$ (left) and partons $\{a, b,c\} = \{q,
\bar{q}, g\}$. The central blob describes the hard scattering cross section
$\hat{\sigma}$. 
The peripheral blobs describe the non-perturbative objects: parton
distributions of the nucleon, $ f_b^N$, and of the photon,
$f_{\protect\rule{0pt}{1.2ex}a}^{\gamma^*}$, and the fragmentation functions of
the produced hadron, $D_c^h$.}
 \label{fig:subprocesses}
\end{figure}

In this Letter, we present a new analysis of COMPASS data for single-inclusive
hadron quasi-real photoproduction at high $p_T$~\footnote{Note that also
inclusive quasi-real photoproduction of hadron pairs can be
considered~\cite{Hendlmeier:2008uw}.}, which differs from  our previous
analysis in that all measured hadrons within a given $p_T$ bin are included in
the analysis, and not only the hadron(s) with highest $p_T$. Moreover, the
interpretation of the results is based on a collinear pQCD
framework that was developed up to NLO~\cite{Jager:2005uf}, the basic concept
being the application of the factorisation theorem to calculate the cross
section of single-inclusive hadron production. The authors of
Ref.~\cite{Jager:2005uf} discuss the sensitivity of COMPASS data to
$\Delta{g(x_g)}$ in terms of contributions from ``direct-photon", $\gamma^*{g}
\rightarrow q\bar{q}$ (Photon Gluon Fusion), and from ``resolved-photon"
subprocesses, $qg$ and $gg$, where the photon acts as a source of partons.
Similarly, they consider direct $\gamma^*{q} \rightarrow qg$ (QCD
Compton) as well as resolved $qq$ and $gq$ subprocesses for the background.
These contributions to the cross section are represented schematically in
\figref{fig:subprocesses}.  In the framework of collinear fragmentation,
photo-absorption on quarks, $\gamma^*{q} \rightarrow q$, is not contributing to
high $p_T$ hadron production.

In order to gain confidence in the applicability of this pQCD framework to
single-hadron production with longitudinally polarised beam and target,  an
important step is to compare predictions of this model to measurements with
beam and target unpolarised, for which the PDFs are well known. While good
agreement was found by RHIC experiments on the production of high-$p_T$ hadrons
in $pp$ collisions at $\sqrt{s} \simeq
200~\GeV$~\cite{Adare:2007dg,Abelev:2009pb}, complications arise when hard
scattering subprocesses are probed in the ``threshold'' regime, in which large
logarithmic contributions from soft and collinear gluons play a significant
role~\cite{deFlorian:2013taa}. Such contributions become dominant at the
COMPASS centre-of-mass energy of $\sqrt{s} \simeq $~18~\GeV. When taken into
account by a technique known as ``threshold resummation'' at next-to-leading
logarithm (NLL)~\cite{deFlorian:2013taa}, the calculations reproduce the
COMPASS cross section measurements~\cite{Adolph:2012nm} within theoretical
uncertainties.

In this Letter, we analyse the quasi-real photoproduction data collected by
COMPASS  from 2002 to 2011 on longitudinally polarised deuteron and proton
targets. In \secref{sec:ExperimentalSetup}, we give a brief description of the
experimental setup, and details on the data selection can be found in
\secref{sec:DataSelection}. The procedure for the asymmetry determination is
described in \secref{sec:asymmetries_and_systematics}. In \secref{sec:Results},
we present the corresponding double spin asymmetries for single-inclusive
hadron production as a function of  their transverse momenta $p_T$. These
asymmetries are compared to
calculations that were performed using the code of Ref.~\cite{Jager:2005uf},
which does not include the resummation of threshold logarithms.

\section{Experimental Setup}
\label{sec:ExperimentalSetup}

The measurements were performed with the COMPASS setup using positive muons
from the M2 beam line of the CERN SPS. A detailed description of the
experimental setup can be found in Ref.~\cite{Abbon:2007pq}, with updates valid
since 2006 described in Ref.~\cite{Abbon201569}. The muon beam had a nominal
momentum of $160$~\GeVoverc, except for 2011 where the momentum was
$200$~\GeVoverc . On average, its momentum spread was $5\,\%$ and its
polarisation was $P_b \approx \val{0.8}$. Momentum and trajectory of incident
muons were measured by a set of scintillator hodoscopes, scintillating fibre
and silicon microstrip detectors. The beam was scattered off a solid state
deuterated lithium ($\mathrm{^6LiD}$) target from 2002 to 2006 and off an
ammonia ($\mathrm{NH_3}$) target in 2007 and 2011, providing longitudinally
polarised deuterons and protons, respectively. The target material was placed
inside a large aperture superconducting solenoid, and by dynamic nuclear
polarisation it was polarised to a value of $P_t \approx \val{0.5}$ for
${\mathrm{^6LiD}}$ and $P_t \approx \val{0.85}$ for ${\mathrm{NH_3}}$. Until
2004, the target material was contained in two contiguous
$\val{60}{\mathrm{cm}}$ long cells that were oppositely polarised. From 2006
onwards, three contiguous target cells of length $\val{30}{\mathrm{cm}}$,
$\val{60}{\mathrm{cm}}$ and $\val{30}{\mathrm{cm}}$ were used to minimise
systematic effects, with the polarisation in the outer cells being opposite to
that in the central one. The direction of the target polarisation was regularly
flipped by reversing the solenoid field to compensate for acceptance
differences between the different target cells. At least once per year, the
direction of the polarisation was reversed relative to that of the solenoid
field. The dilution factor $f$, which accounts for the presence of
unpolarisable material, amounts to typically 0.4 for the deuterated lithium
target and to 0.16 for the ammonia one. It is calculated as the ratio of the
cross section on polarisable deuteron or proton to that on all target nuclei,
corrected for unpolarised $x$ and $y$ dependent electromagnetic radiative
effects~\cite{Alexakhin2007330}, where $x$ is the Bjorken scaling variable and
$y$ the relative muon energy transfer. No further radiative effects are taken
into account. The momenta and angles of scattered muons and produced hadrons
were measured in the two-stage open forward spectrometer, where each stage
includes a dipole magnet with upstream and downstream tracking detectors.

\section{Data Selection}
\label{sec:DataSelection}
In order to be selected, an event must have an interaction vertex that contains
both incoming and scattered muons and at least one hadron candidate track. The
measured beam momentum is required to be in a $\pm 20$~GeV interval around the
nominal value ($\pm 15$~GeV in 2011). In order to equalise the flux through
each target cell, the extrapolated beam track is required to pass all target
cells. Cuts on the position of the vertex allow the selection of the target
cell, in which the scattering occurred. Only events with photon virtuality
$Q^2<1\  (\GeVoverc)^2$ are accepted. This kinematic region is referred to in
the following as quasi-real photoproduction region. In addition, $y$ is
required to be within 0.1 and 0.9, where the lower limit removes events that
are difficult to reconstruct and the upper limit removes the region where
electromagnetic radiative effects are large. These kinematic cuts result in a
range of $10^{-5} < x < 0.02$ and a minimum mass squared of the hadronic final
state, $W^2$, of 25 $(\GeVovercsq)^2$. The hadron candidate track must have
$p_T > 0.7\ \GeVoverc$. The fraction $z$ of the virtual photon energy carried
by the hadron is required to be in the range $0.2 < z < 0.8$, where the lower
limit is imposed to suppress the contribution from target remnant hadronisation
and the upper limit to reject badly reconstructed hadrons. The angle between
the direction of the hadron and that of the virtual photon is restricted to be
in the range $10\,\mathrm{mrad} < \theta < 120\,\mathrm{mrad}$, which
corresponds to $2.4 > \eta > -0.1$, where $\eta$ is the pseudo-rapidity in the
$\gamma^*N$ centre-of-mass system. After all selections, the final sample
consists of 140 million events for the deuteron target and 105 million for the
proton target.

\section{Asymmetry Calculation}
\label{sec:asymmetries_and_systematics}
The double-spin asymmetry of the cross sections for single hadron quasi-real
photoproduction is defined as
$ A_{LL} =(\sigma^{\leftrightarrows} - \sigma^{\leftleftarrows}) /
(\sigma^{\leftrightarrows} + \sigma^{\leftleftarrows}) = \Delta \sigma /
\sigma$,
where the symbols $\leftrightarrows$ and $\leftleftarrows$ denote anti-parallel
and parallel spin directions, respectively, of the incident muon and the target
deuteron or proton. This asymmetry is evaluated using the same method as in our
previous analyses~\cite{Adolph:2012vj}. The number of hadrons produced in a
target cell is related to $ A_{LL}$ and to the spin independent cross section
$\sigma= \sigma^{\leftrightarrows} + \sigma^{\leftleftarrows}$: $N_i= a_i
{\phi}_i n_i \sigma (1+ f_i P_b P_{t_i} A_{LL})$, where $i=u_1,d_1,u_2,d_2$. A
target cell (\textit{u} or \textit{d}) with a given direction of the target
polarisation (1 or 2) has the acceptance $a_i$, the incoming muon flux $\phi_i$
and the number of target nucleons $n_i$. For the two-cell target, \textit{u}
and \textit{d} denotes upstream and downstream cell, respectively, while for
the three-cell target, \textit{u} denotes the sum of the outer cells and d the
central cell. The asymmetry $A_{LL}$ is extracted from the second order
equation that is obtained from the quantity $(N_{u_1} \cdot N_{d_2}) / (N_{d_1}
\cdot N_{u_2})$. In this relation, fluxes and acceptances cancel, provided that
the ratio of acceptances of the two sets of cells is equal for the two
orientations of the solenoid field.

In order to minimise statistical uncertainties, all quantities entering the
asymmetry are calculated for each hadron using a weight factor $w_i = f_i
P_b$~\cite{PhysRevD.56.5330}. The muon beam polarisation $P_b$ is obtained from
a parametrisation as a function of the beam momentum. The target polarisation
$P_{t_i}$ is not included in the weight $w_i$ as it changes with time and could
generate false asymmetries. In order to reduce systematic uncertainties, data
are grouped into periods that are close in time and hence have the same
detector conditions, and the weighted average over all periods is taken. The
asymmetries determined for a given target from data taken in different years
were found to be consistent and hence combined. The asymmetries are obtained
for both positive and negative unidentified hadrons in bins of $p_T$ in the
range $0.7$~\GeVoverc \ to $4$~\GeVoverc \ and in bins of $\eta$ in the range
$-0.1$ to $2.4$ in order to facilitate a detailed comparison to theory (see
\secref{sec:Results}). The data for $p_T < 1.0$~\GeVoverc \ are only used to
investigate systematic uncertainties as the pQCD framework is commonly applied
only for hard scales $\val{\mu^{2} \simeq p_T^{2} \ge 1.0}{(\GeVoverc)^2}$, and
they are shown greyed out in all the figures where they appear in.

The systematic uncertainties on $A_{LL}$ are calculated as the square root of
the sum of squares of multiplicative and additive contributions. The
uncertainties on the dilution factor ($\approx 5\%$),  the beam ($\approx 5\%$)
and the target ($\approx 5\%$) polarisations contribute to a total of $\approx
8\%$ of multiplicative uncertainties, \textit{i.e.} those being proportional to
the asymmetry itself. Additive contributions originate from fluctuations of the
detector performance, which may lead to false asymmetries. Their possible
occurrence is investigated by dividing the data sample into different subsets.
Asymmetries calculated with hadrons detected in left and right (top and bottom)
parts of the spectrometers are found to be compatible within statistical
uncertainties, as well as those for the two relative orientations of the
solenoid field and the target spin vectors. No systematic uncertainty is thus
attributed to these effects. Possible false asymmetries  between data sets
having the same polarisation states are also found to be compatible with zero.
For each $p_T$ bin, the statistical distribution of the asymmetries calculated
by time periods closely follows a normal distribution. The observed deviations
from a Gaussian allow us to quantify  the level of overall additive systematic
uncertainties as a fraction of the statistical ones, which on average amounts
to about one half. These additive systematic uncertainties largely dominate
over the multiplicative ones.

\section{Results and interpretation}
\label{sec:Results}

\begin{figure}[htbp]
 \includegraphics[width=1.\linewidth]{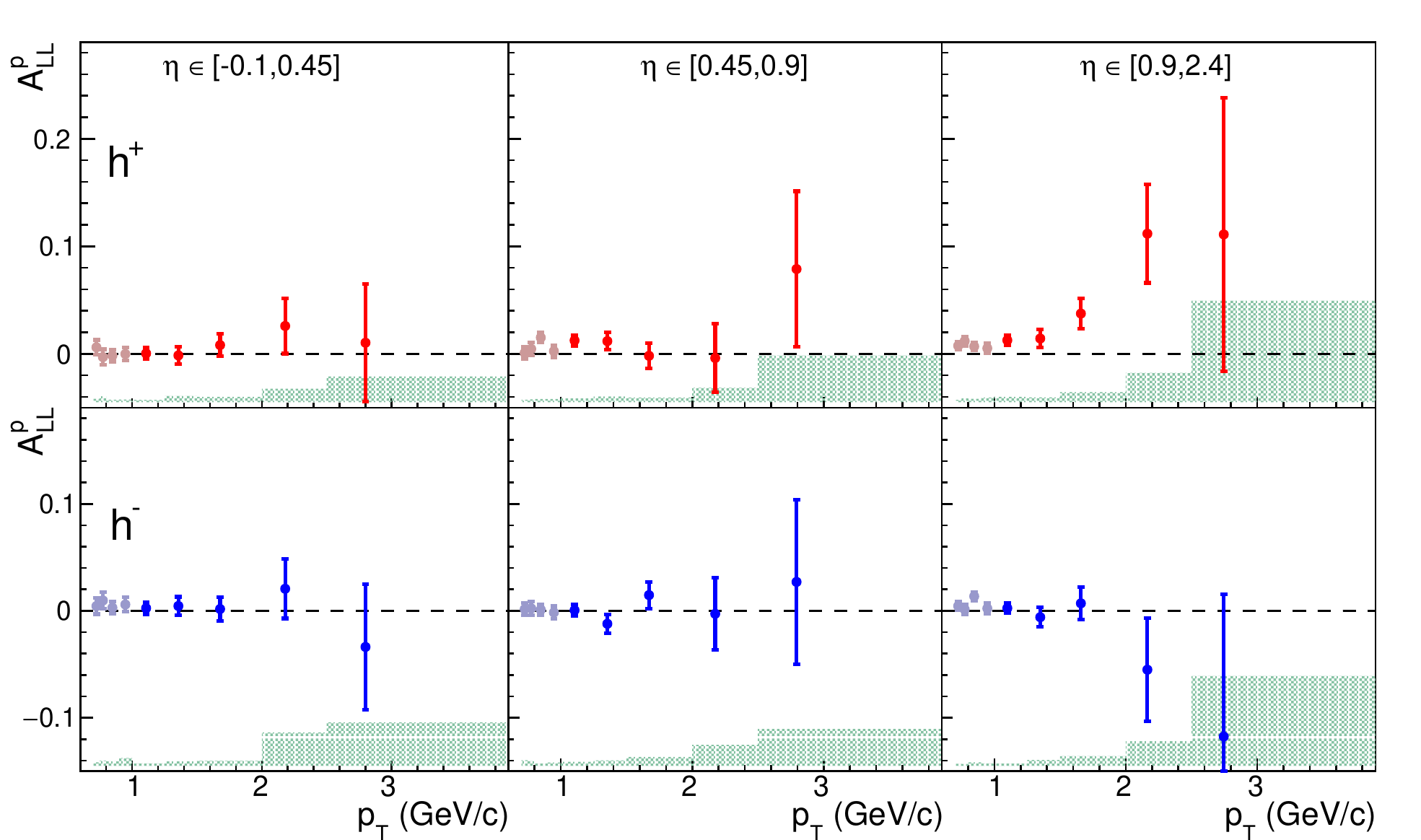}
 \caption{The asymmetry $A_{LL}$ as a function of $p_{T}$ for charged hadron
quasi-real photoproduction on the proton for three rapidity bins. The bands at
the bottom indicate the systematic uncertainties, which are dominated by time
dependent fluctuations (see \secref{sec:asymmetries_and_systematics}.) Top:
positive hadron production; bottom: negative hadron production. }
 \label{all_prot}
\end{figure}

\begin{figure}[htbp]
 \includegraphics[width=1.\linewidth]{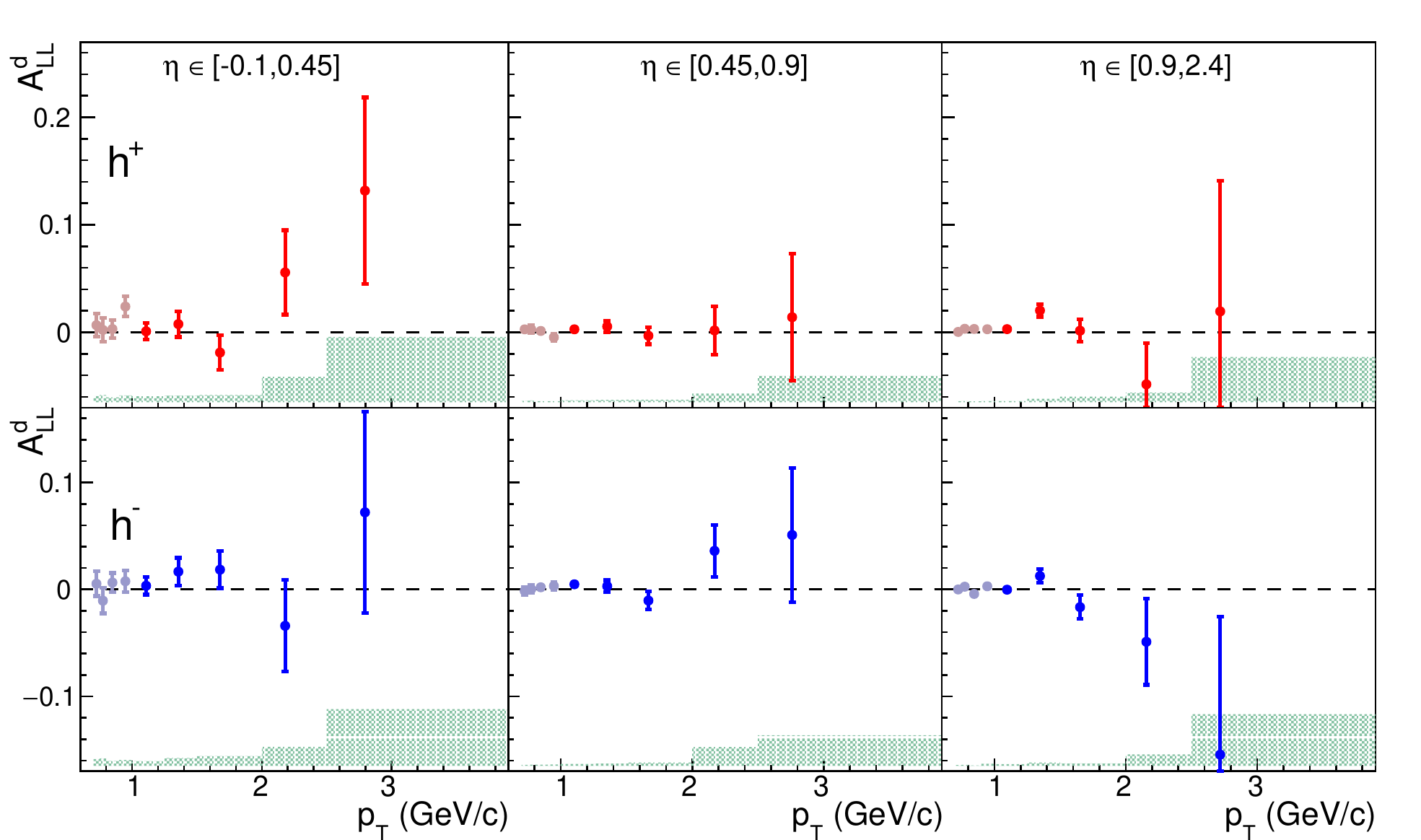}
  \caption{Same as \figref{all_prot}, but for the deuteron.}
 \label{all_deut}
\end{figure}

The final asymmetries are calculated using all data accumulated with the
deuteron target in the years 2002 to 2006 and with the proton target in the
years  2007 and 2011. Their $p_T$-dependence in three rapidity bins spanning
the full interval $-0.1 < \eta < 2.4$ ($[-0.1,0.45]$,  $[0.45,0.9]$, and
$[0.9,2.4]$) is shown in \figref{all_prot} and \figref{all_deut} for each
target type and hadron charge.

We compare our asymmetries with theoretical calculations at NLO without
threshold resummation based on the framework described in
Ref.~\cite{Jager:2005uf} and summarised in the following. Using the code of
Ref.~\cite{Jager:2005uf}, the asymmetries are computed in bins of $p_T$ and
$\eta$ as the ratio of polarised to unpolarised hadron cross sections, where a
cross section is a convolution of the ``muon--parton distribution function"
$f_a^{\mu}$, the nucleon PDFs $f_b^N$, the perturbative partonic cross sections
$\hat{\sigma}_{a+b\rightarrow c+X}$, and the fragmentation functions (FF)
$D_c^h$:

{\small \begin{align}
 A_{LL}(p_T,\eta)
  = \frac{\diff \Delta \sigma^h}{\diff \sigma^h}(p_T,\eta)
  = \frac{\sum _{a,b,c} \Delta f_a^{\mu} \otimes \Delta f_b^N \otimes \diff \Delta \hat{\sigma}_{a+b\rightarrow c+X} \otimes D_c^h}
         {\sum _{a,b,c} f_a^{\mu} \otimes f_b^N \otimes \diff \hat{\sigma}_{a+b\rightarrow c+X} \otimes D_c^h}\ .
 \label{eq:th_def_asym}
\end{align}}

Here and below, spin-dependent quantities are denoted by the symbol $\Delta$
and will be referred to as polarised ones in the rest of the Letter
(spin-independent ones as unpolarised). The processes involved in
\eqref{eq:th_def_asym} can be classified into ``direct" ones that are initiated
by a quasi-real photon and ``resolved" ones that are initiated by its
fluctuation into partons. This classification is denoted by the subscript $a$
(see \figref{fig:subprocesses}). For direct processes, subscript $a$ refers to
$\gamma^*$, and $(\Delta)f_{\gamma^*}^\mu$ is the probability for a muon to
emit a quasi-real photon. For resolved processes, subscript $a$ refers to $q$,
$\bar{q}$ 
or $g$, and $(\Delta)f_{a}^\mu$ is the convolution of this probability with a
non-perturbative parton distribution of the photon,
($\Delta$)$f_{\protect\rule{0pt}{1.2ex}a}^{\gamma^*}$. The polarised version of
the latter is not known experimentally and hence taken to range between the
positive and negative magnitude of the unpolarised one. This induces a small
uncertainty in the theoretical calculations.\\

\begin{figure}[htbp]
\begin{center}
 \includegraphics[width=.49\linewidth]{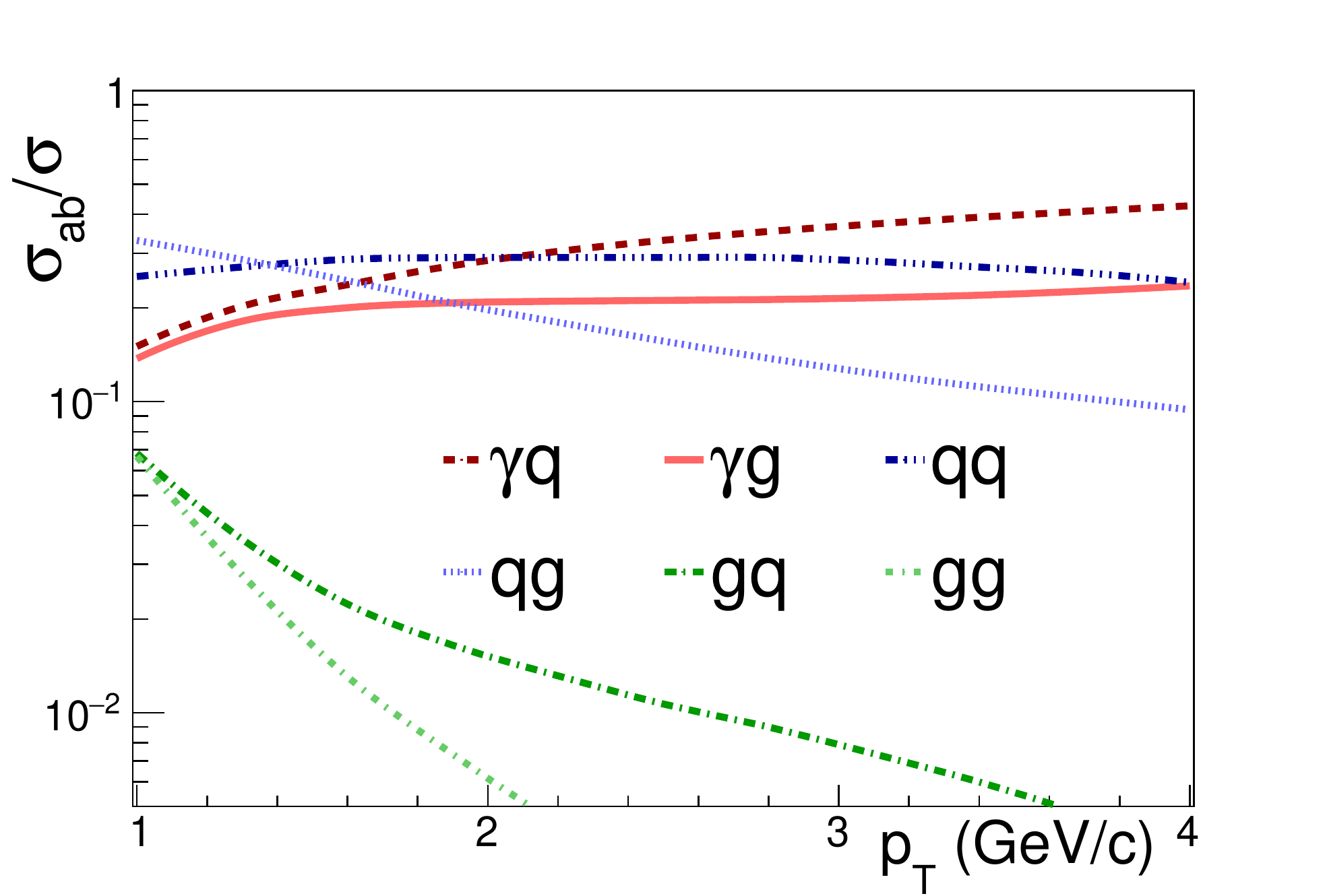}
 \includegraphics[width=.49\linewidth]{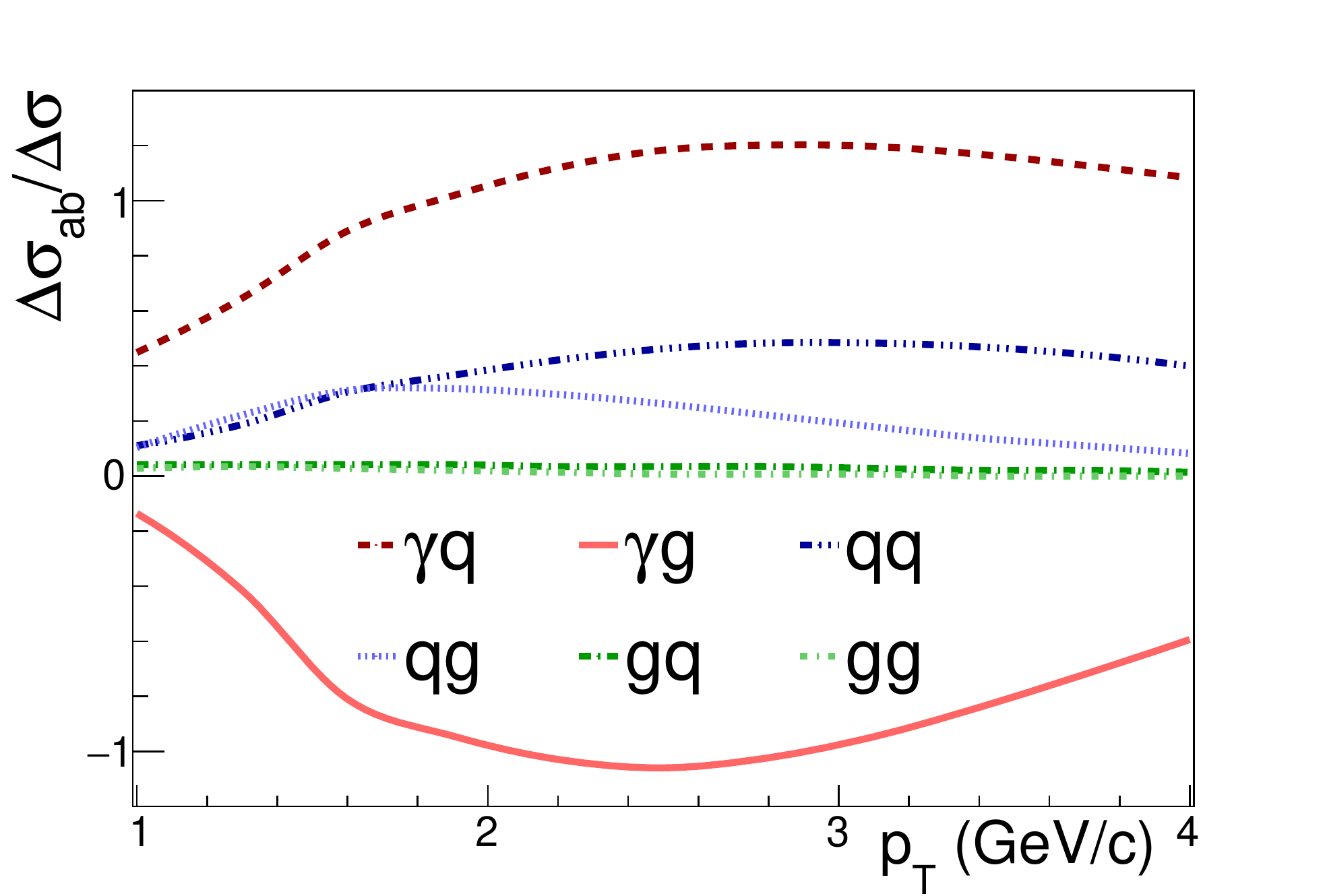}
 \caption{\label{fig:ratio_process}
Contributions of the six subprocesses $a+b \rightarrow c+X$ described
in \secref{sec:Introduction} to the full NLO unpolarised (left) and polarised
(right) photoproduction cross sections for a deuteron target. The polarised
cross sections are computed using the polarised PDF set of
Ref.~\cite{deFlorian:2014yva}.}
 \end{center}
\end{figure}

The values of the asymmetries are computed here using the following input
distributions: the unpolarised parton distributions of the photon
$f_{\protect\rule{0pt}{1.2ex}a}^{\gamma^*}$ from GRS\cite{grs}, the unpolarised
nucleon PDFs $f_b^N$ from 
CTEQ65\cite{Pumplin:2002vw}, the three polarised PDF sets from
GRSV\cite{Gluck2001} as in Ref.~\cite{Jager:2005uf} (the ``standard" set and the
two sets for ``maximum" [$\Delta g(x)=g(x)$] and ``minimum" [$\Delta g(x)=-g(x)$]
gluon distribution functions at input scale), as well as the most recent
polarised PDF set DSSV14 from Ref.~\cite{deFlorian:2014yva}.  For the polarised
PDF sets used, the integration over the range $0.05 <  x_g < 0.2$,
which is characteristic for the kinematic coverage of COMPASS in the gluon
momentum fraction, yields the following ``truncated" values of $\Delta G$ at a scale of 
$3\ (\GeVoverc)^2$ : $\Delta G_{GRSV_{min}} \approx -0.6$, $\Delta
G_{DSSV14} \approx 0.1$, 
$\Delta G_{GRSV_{std}} \approx 0.2$, $\Delta G_{GRSV_{max}} \approx 0.7$. The other 
inputs that we changed with respect to Ref.~\cite{Jager:2005uf} are the 
fragmentation functions $D_c^h$, for which we use the most recent
parton-to-pion fragmentation set of Ref.~\cite{PhysRevD.91.014035}, which best 
fits the recent COMPASS pion multiplicities~\cite{PosDIS2013}. We checked, as 
it was  done in Ref.~\cite{Jager:2005uf}, that asymmetries for hadron and pion 
production are almost indistinguishable, so that it is safe to compare our 
experimental data to theoretical asymmetries computed with a parton-to-pion FF 
set.
The fractions of the total unpolarised (respectively polarised) cross section
for the various individual
subprocesses are shown in \figref{fig:ratio_process} as a function
of $p_T$.  
Although the unpolarised cross sections for processes involving gluons from the nucleon are
not the dominant ones, the polarised cross section for the $\gamma{g}$ subprocess is large in
magnitude for hadron production at high $p_T$, which makes the study of such 
asymmetries relevant in the COMPASS
kinematic region.

\begin{figure}[htbp]
\includegraphics[width=1.\linewidth]{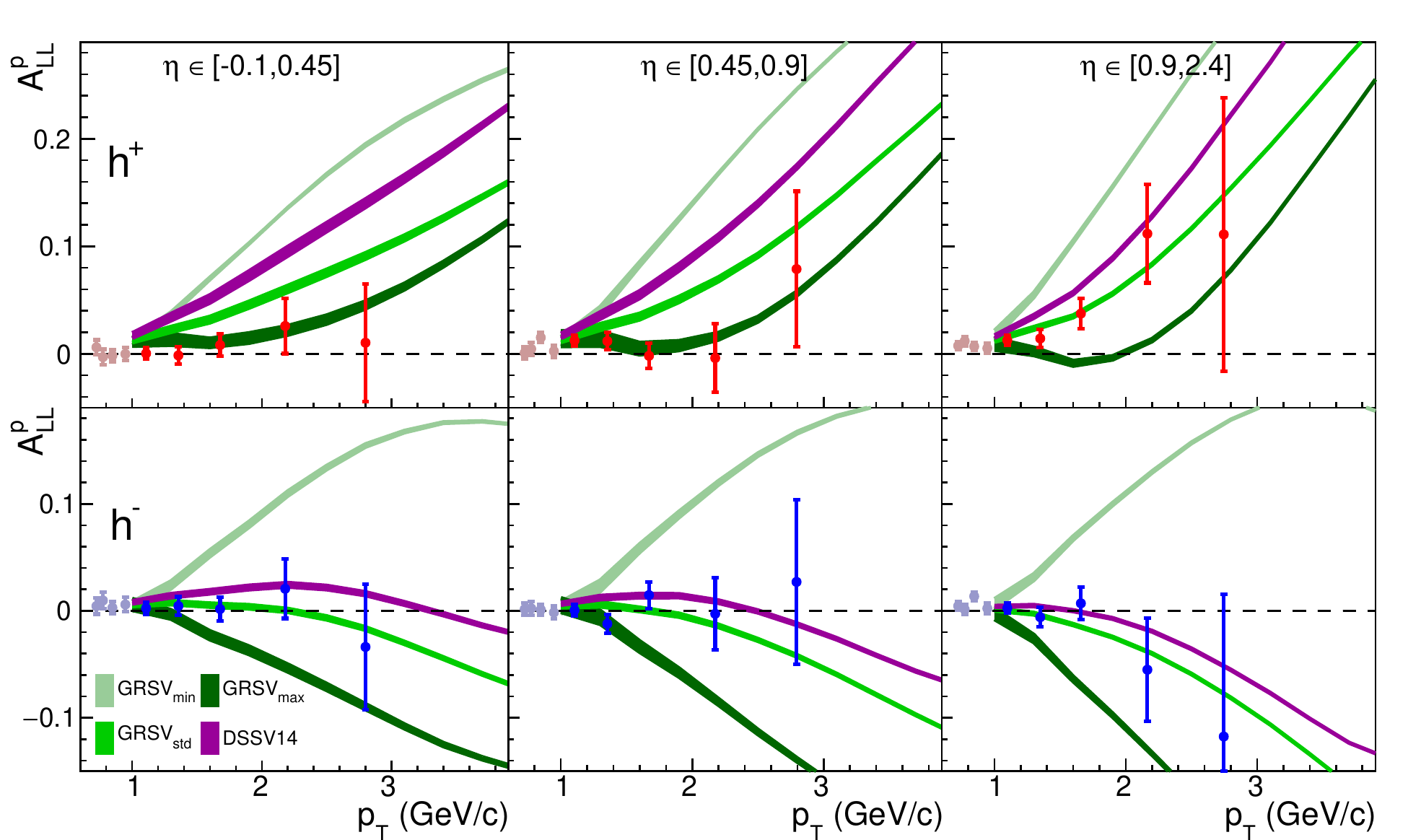}
 \caption{COMPASS asymmetries $A_{LL}$ for a proton target as a function of
$p_{T}$ and in three rapidity bins, compared to NLO calculations based on
Ref.~\cite{Jager:2005uf} for different choices for the polarised PDFs (see
text). Only statistical uncertainties are shown. Error bands 
on the theory curves represent the uncertainties due to the polarised parton
distribution of the photon. Top: positive hadron production; bottom: negative
hadron production.}
 \label{fig:all_prot_theo}
\end{figure}

\begin{figure}[htbp]
\includegraphics[width=1.\linewidth]{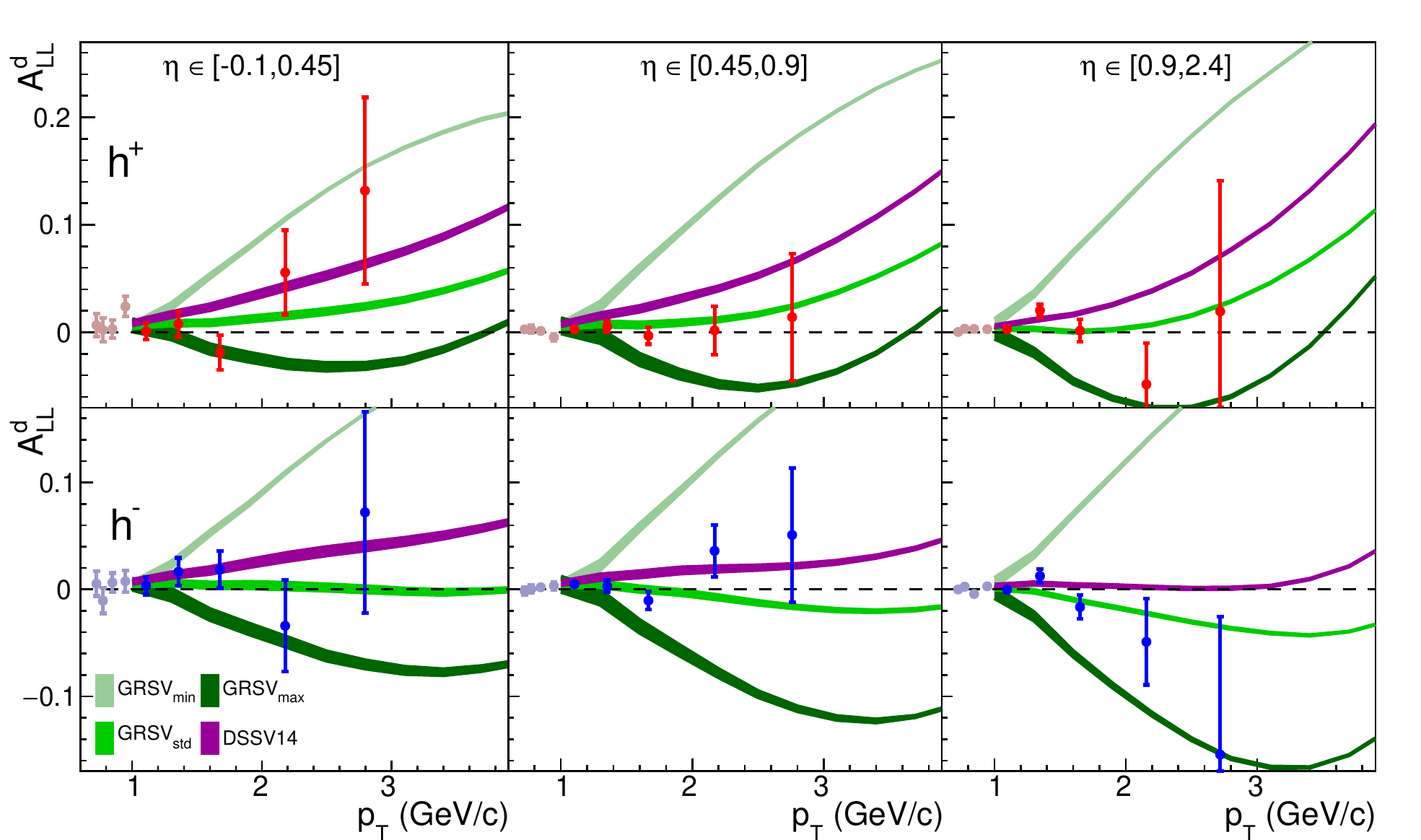}
 \caption{Same as \figref{fig:all_prot_theo}, but for a deuteron target.}
 \label{fig:all_deut_theo}
\end{figure}

The computations of $A_{LL}(p_T)$ are performed at COMPASS kinematics using the
same cuts as in the present data analysis, \textit{i.e.},
$\val{p_T>1}{\GeVoverc}$, $\val{Q^2<1}{(\GeVoverc)^2}$, $0.1<y<0.9$,
$0.2<z<0.8$. For consistency, we verified that when using the same inputs as in
Ref. \cite{Jager:2005uf}, we reproduce the asymmetry calculated there. The
computations are done separately for the production of positive and negative
hadrons, and for three distinct bins in $\eta$: $[-0.1,0.45]$, $[0.45,0.9]$,
and $[0.9,2.4]$. The results of these computations are compared to the
experimental asymmetries in \figref{fig:all_prot_theo} and 
\figref{fig:all_deut_theo}.

The data are seen to be consistent with the NLO calculations of
Ref.~\cite{Jager:2005uf} using the most recent polarised
PDF~\cite{deFlorian:2014yva} and FF~\cite{PhysRevD.91.014035} sets, except for
positive hadron production from the proton in the rapidity range $-0.1 < \eta <
0.9$. Our data are also compared in \figref{fig:all_prot_theo} and
\figref{fig:all_deut_theo} to calculations using earlier GRSV polarised
PDFs~\cite{Gluck2001} to give an impression of their sensitivity to $\Delta G$,
which seems enhanced at higher values of $\eta$.
A possible reason for the discrepancy seen at low $\eta$ in positive hadron
production from the proton may be that the model 
calculations had to be done without threshold resummation at NLL, as the
formalism for the polarised case is not yet fully available. However, 
contrary to the unpolarised case where data can only be described by pQCD if 
threshold resummation is included~\cite{deFlorian:2013taa}\footnote{In 
Ref.~\cite{Adolph:2012nm}, the validity of the 
formalism~\cite{deFlorian:2013taa} was verified in the full $\eta$ and $p_T$
range covered by COMPASS kinematics. This statement does not suffer from
slightly larger $y$ and $Q^2$ cuts used in the present analysis to enhance
statistics, as we checked that the asymmetries obtained for the two sets of
cuts are compatible within statistical uncertainties.}, spin asymmetries are
expected to be less affected~\cite{PhysRevD.87.094021}.
A first estimation of
the impact of threshold resummation on our double spin asymmetries including so
far only the direct processes~\cite{Vogelsang} indicates a substantial dilution
of the asymmetries, which may explain part of the discrepancy between
experiment and theory for positive hadron production on the proton at low
values of $\eta$.

\section{Summary}
\label{sec:Summary}
In summary, we have presented in this Letter a new analysis of COMPASS data on
polarised single-inclusive hadron quasi-real photoproduction, which in
principle is well  suited for an extraction of the gluon polarisation $\Delta
G$ in the framework of collinear pQCD. Results for the longitudinal spin
asymmetry $A_{LL}(p_T)$ on polarised protons and deuterons are given separately
for positively and negatively charged hadrons, and in three rapidity bins. They
are compared to theoretical calculations at NLO without threshold resummation
and overall agreement is found with the calculations based on earlier
GRSV$_{std}$ and recent DSSV14 polarised PDF sets, and using the most recent FF
set. Nevertheless, calculations including full threshold resummation at NLL are
needed before a meaningful result on $\Delta G$ can be extracted quantitatively
from our data.

\section*{Acknowledgements}
\label{sec:Acknowledgements}
We thank Werner Vogelsang and Marco Stratmann for many useful discussions and
for providing us the codes for the NLO pQCD calculation. We gratefully
acknowledge the support of the CERN management and staff and the skill and
effort of the technicians of our collaborating institutes. This work was made
possible by the financial support of our funding agencies.

\end{document}